\documentclass[twocolumn,
hidelinks,
superscriptaddress,
floatfix,preprintnumbers]{revtex4-2}
\usepackage{graphics,amssymb,amsmath,epsfig,color}
\usepackage{graphicx}
\usepackage{minitoc}

\usepackage{svg}
\usepackage[colorlinks=true,
            linkcolor=blue,
            urlcolor=blue,
            citecolor=blue]{hyperref}

\begin{document}

\title{Gate-tunable electroresistance in a sliding ferroelectric tunnel junction}

\author{Bozo Vareskic}
\email{bv227@cornell.edu}
\affiliation{Department of Physics, Cornell University, Ithaca, NY, 14853, USA}
\author{Finn G. Kennedy}
\affiliation{Department of Applied and Engineering Physics, Cornell University, Ithaca, NY, 14853, USA}
\author{Takashi Taniguchi}
\affiliation{Research Center for Materials Nanoarchitectonics, National Institute for Materials Science,  1-1 Namiki, Tsukuba 305-0044, Japan}
\author{\\Kenji Watanabe}
\affiliation{Research Center for Electronic and Optical Materials, National Institute for Materials Science, 1-1 Namiki, Tsukuba 305-0044, Japan}
\author{Kenji Yasuda}
\affiliation{Department of Applied and Engineering Physics, Cornell University, Ithaca, NY, 14853, USA}
\author{Daniel C. Ralph}
\email{dcr14@cornell.edu}
\affiliation{Department of Physics, Cornell University, Ithaca, NY, 14853, USA}
\affiliation{Kavli Institute for Nanoscale Science, Cornell University, Ithaca, NY 14853, USA}

\date{\today}
\begin{abstract}
We fabricate and measure electrically-gated tunnel junctions in which the insulating barrier is a sliding van der Waals ferroelectric made from parallel-stacked bilayer hexagonal boron nitride and the electrodes are single-layer graphene. Despite the nominally-symmetric tunnel-junction structure, these devices can exhibit substantial electroresistance upon reversing the ferroelectric polarization. The magnitude and sign of tunneling electroresistance are tunable by bias and gate voltage. We show that this behavior can be understood within a simple tunneling model that takes into account the quantum capacitance of the graphene electrodes, so that the tunneling densities of states in the electrodes are separately modified as a function of bias and gate voltage.

\end{abstract}
\pacs{}
\maketitle

Ferroelectric tunnel junctions (FTJs) are promising candidates for next-generation memory technologies due their capacity for non-volatile memory storage, non-destructive readout, and low write energy \cite{garcia_ferroelectric_2014, park_ferroelectric_2024}. In an FTJ, a ferroelectric layer which serves as an insulating tunnel barrier is sandwiched between two conducting electrodes. In order for the tunneling conductance to be sensitive to the polarization direction, thereby giving a non-zero tunneling electroresistance (TER), the junction structure must not be mirror symmetric.  This has been achieved in previous work by using electrodes with different screening lengths \cite{zhuravlev_giant_2005, wu_high_2020} or by inserting a dielectric spacer between the ferroelectric layer and one electrode \cite{zhuravlev_tunneling_2009,gao_tunnel_2024}. 
Making ferroelectric layers which are simultaneously thin enough to serve as tunnel barriers and stable against depolarization and wear-out processes is challenging with conventional ferroelectric materials, in part because the depolarization field from the bound charges can destabilize ferroelectricity for very thin layers \cite{junquera_critical_2003, kim_critical_2005}.  This has motivated interest in ferroelectrics made from van der Waals materials which show no critical thickness for ferroelectric order \cite{chang_discovery_2016, liu_room-temperature_2016, zhou_out--plane_2017, cui_intercorrelated_2018, yuan_room-temperature_2019, song_evidence_2022, wang_towards_2023}. Some van der Waals materials, e.g., transition metal dichalcogenides \cite{fei_ferroelectric_2018, jindal_coupled_2023, wang_interfacial_2022} and hexagonal boron nitride (hBN) \cite{li_binary_2017, yasuda_stacking-engineered_2021, woods_charge-polarized_2021, vizner_stern_interfacial_2021} provide a novel form of ferroelectricity, sliding ferroelectricity, in which switching is achieved by the relative sliding motion of entire van der Waals layers. This sliding mechanism combined with the atomically pristine nature of van der Waals layers can provide higher endurance compared to non-sliding ferroelectrics \cite{bian_developing_2024, yasuda_ultrafast_2024}. 

Here, we demonstrate another potential advantage of assembling ferroelectric tunnel junctions from van der Waals materials -- they allow the added functionality of making the TER tunable by means of electrical gating.  We show that a nominally mirror-symmetric FTJ with the structure graphene electrode/bilayer hBN/graphene electrode can nevertheless still provide a substantial TER signal upon switching of the ferroelectric bilayer hBN, because the graphene electrodes can be electrically gated so that they have different densities of states for tunneling. The resulting TER exhibits a controllable pattern with multiple sign changes as a function of gate voltage ($V_G$) and bias voltage across the tunnel junction ($V$).  We explain how this pattern can be understood by taking into account the quantum capacitance of the graphene electrodes in order to track how the electron chemical potentials in the electrodes shift relative to their Dirac points as a function of $V_G$ and $V$.

\begin{figure}[htbp]
  \centering
  \includegraphics[scale = 1.1]{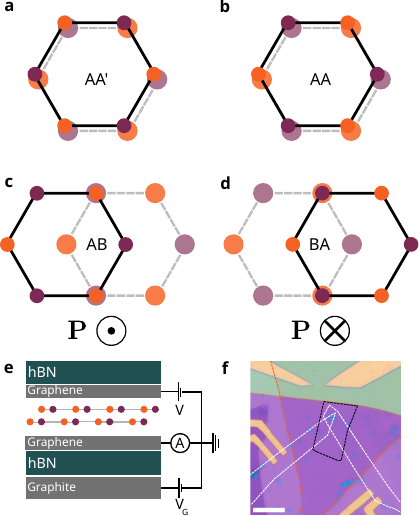}
  \caption{\textbf{Sliding ferroelectric tunneling transistor.}  (a)-(d) Bilayer hBN allotropes. Purple (orange) circles represent boron (nitrogen) atoms while the smaller and darker (fainter and larger) circles correspond to the top (bottom) layer. (a) shows the anti-parallel AA' structure of bilayer hBN as exfoliated from bulk where the atoms of one layer fully eclipse the atoms of the layer below. (b) depicts the unstable AA parallel stacking order. The two layers are slightly offset for clarity in (a) and (b). (c)-(d) show the two ferroelectric allotropes AB and BA where the ferroelectric order points out of or into the page. (e) Diagram of device structure (not to scale). For all measurements the bottom graphene layer is grounded. (f) Micrograph of device. The dashed white, black, and orange lines outline the graphene electrodes, P-BBN, and graphite gate electrode, respectively. Scale bar is 10 $\mu$m. 
  }
  \label{fig_1}
\end{figure}

\section{Device Structure}
For the ferroelectric tunnel barrier in our devices we use parallel-stacked bilayer boron nitride (P-BBN). When simply exfoliated from bulk hBN, native bilayer hBN (Fig.\ \ref{fig_1}(a)) exhibits AA’ stacking where the top layer is rotated 180$^\circ$ (antiparallel) relative to the bottom layer, in which case the boron (nitrogen) atoms of the top layer fully eclipse the nitrogen (boron) atoms of the bottom layer, so that there is no net ferroelectricity. Parallel bilayer boron nitride can be obtained by tearing a monolayer hBN flake and placing one half of it on top of the other while maintaining the relative angular alignment between the two layers. Fig.\ \ref{fig_1}(b) depicts P-BBN in the AA stacking configuration where the boron (nitrogen) atoms of the top layer eclipse the boron (nitrogen) atoms of the bottom layer. The AA stacking configuration is energetically unfavorable however, and the bilayer will transition to one of two degenerate ferroelectric allotropes, AB (Fig.\ \ref{fig_1}(c)) or BA (Fig.\ \ref{fig_1}(d)), which host opposite out-of-plane electrical polarizations. P-BBN can be switched between the AB and BA phase by applying an out of plane electric field \cite{vizner_stern_interfacial_2021, yasuda_stacking-engineered_2021}. This switching is driven by the sliding motion of one entire atomic layer relative to the other by a distance of the B-N bond length. 

We incorporate P-BBN into an FTJ by using mechanical stacking to make heterostructures in which the P-BBN is sandwiched between two flakes of monolayer graphene which serve as the electrodes. The top of the heterostructure is encapsulated by a 75 nm thick hBN dielectric layer while the bottom is encapsulated by a 70 nm hBN dielectric layer and a 2.6 nm graphite gate electrode as shown in Fig.\ \ref{fig_1}(e) and (f). The inclusion of a gate allows us to tune the electron chemical potentials of the junction electrodes. For details on the device fabrication, see Supplementary Information Section S1. We measure the junction differential conductance $dI/dV$ as a function of $V$ at different fixed values of $V_G$ by applying a small (1 mV) AC voltage on top of the DC bias and measuring the resulting AC current with a lock-in amplifier. All measurements are performed at $T = 4.2$ K.

\section{Nature of tunneling}

\begin{figure*}[htbp]
  \centering
  \includegraphics[scale = 1]{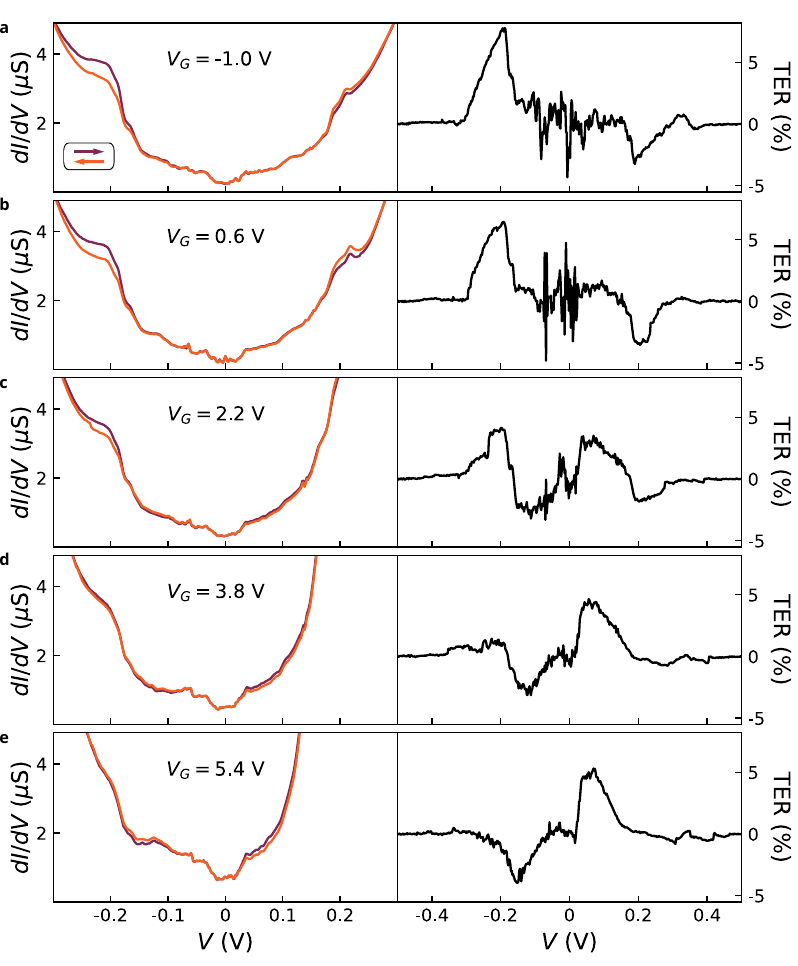}
  \caption{  \textbf{Evolution of TER with gate voltage.} (a) Left panel: Tunneling conductance as a function of bias voltage at $V_G = -1.0$ V and 4.2 K for a forward (purple) and backward (orange) sweep. Right panel: TER of the left panel. (b)-(d) are the same as (a) but are measured at $V_G = 0.6$ V, $2.2$ V, $3.8$ V, and 5.4 V respectively. 
  }
  \label{fig_2}
\end{figure*}

We investigate the gate dependence of the TER by performing forward and backward sweeps of $V$ at constant $V_G$. The left panel of Fig.\ \ref{fig_2}(a) shows the differential conductance $dI/dV$ at $V_G = -1.0$ V as $V$ is swept forward (purple) from $V = -0.5$ V to $V = 0.5$ V and backward (orange) from $V = 0.5$ V to $V = -0.5$ V. Both forward and backwards sweeps reach a minimum of $dI/dV \sim$ 240 nS when $|V| < 25$ mV and then $dI/dV$ rises as $V$ increases in magnitude. The increasing conductance is accompanied by step-like jumps positioned symmetrically in $V$. The evolution of the differential conductance with $V_G$ is illustrated in the left-hand panels of Fig.\ \ref{fig_2}(b)-(e).

\begin{figure*}[htbp]
  \centering
  \includegraphics[scale = 1.2]{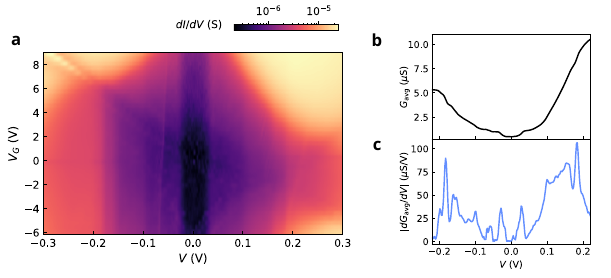}
  \caption{\textbf{Analysis of inelastic electron tunneling.} (a) Junction conductance as a function of $V$ and $V_G$. $V$ is swept from negative to positive. (b) Average conductance versus $V$. The average is taken over scans across all the measured gate voltages in ordered to emphasize the step-like features. (c) Derivative of average conductance. Step-like features in $G_{\text{avg}}$ line up with peaks in $|dG_{\text{avg}}/dV|$. The positions of these peaks correspond to threshold energies for the onset of inelastic tunneling. }
  \label{fig_3}
\end{figure*}

The non-Ohmic behavior of $dI/dV$ confirms that the P-BBN acts a good tunneling barrier between the conducting graphene layers.  The step-like features reveal that the majority of the tunneling current is due to phonon-assisted inelastic tunneling \cite{amet_tunneling_2012, jung_vibrational_2015, chandni_signatures_2016, vdovin_phonon-assisted_2016}. 
To highlight the importance of inelastic tunneling, in Figure \ref{fig_3}(a) we show a color plot of the tunneling conductance (on a logarithmic scale) as a function of $V$ and $V_G$ on a forward bias-voltage sweep. The conductance displays prominent vertical features that correspond to step-like jumps in the conductance which persist through all gate voltages and are symmetric in $V$. 
Figure \ref{fig_3}(b) shows the result if we average across all gate voltages to obtain $G_{\text{avg}}$, and Fig.~\ref{fig_3}(c) shows the derivative $|dG_{\text{avg}}/dV|$. The step-like jumps in $G_{\text{avg}}$ near $V = \pm28$ meV, $\pm 100$ meV, $\pm 152$, and $\pm 184$ meV lead to peaks in $|dG_{\text{avg}}/dV|$ at these voltages.  Peaks near $V = \pm 28$ meV in previous studies of graphene/boron nitride/graphene tunnel junction have been attributed to an out of plane acoustic (ZA) phonon mode \cite{jung_vibrational_2015, vdovin_phonon-assisted_2016}.  In crystalline graphene/hBN/graphene devices like ours in which the graphene lattices are not aligned (so that the low-energy $K$ and $K'$ points in the two graphene layers are also mis-aligned), elastic tunneling is expected to be suppressed in the range of voltages we examine due to the need to conserve crystal momentum in elastic tunneling \cite{vdovin_phonon-assisted_2016, perebeinos_phonon-mediated_2012}. The small non-zero conductance we observe below the first inelastic threshold is therefore likely due to disorder or impurities.

\section{Gate-tunable tunneling electroresistance}
Our main focus is the hysteresis in $dI/dV$ relative to the two sweep directions of the bias voltage, which is indicative of switching of ferroelectric order in the barrier layer.  We define a TER as 
\begin{equation}\label{ter}
    \text{TER} = 100 \times \frac{G_\text{f}-G_\text{b}}{G_\text{f}+G_\text{b}}
\end{equation}
where $G_{\text{f}(\text{b})}$ is the differential conductance of the forward (backward) sweep. The right panels of Fig.\ \ref{fig_2} show the TER at different values of $V_G$. The sign of the measured TER depends on both $V_G$ and $V$. At $V_G = -1.0$ V, the TER reaches a maximum value of $7.8\%$ at $V  = -0.19$ V and changes sign at positive bias voltage with a negative peak of $-3.2\%$ at $V = 0.21$ V. For low biases $|V| < 0.1$ V, the TER is small relative to the peak values and noisy. The TER vanishes above approximately $V = 0.4$ V. This gives an estimate of the coercive field $E_c$ as $E_c \approx 0.44$ V/nm. The coercive field however can vary with gate voltage due to incomplete screening of the electric field from the gate electrode \cite{yasuda_stacking-engineered_2021}.

\begin{figure*}[htbp]
  \centering
  \includegraphics[scale = 1.1]{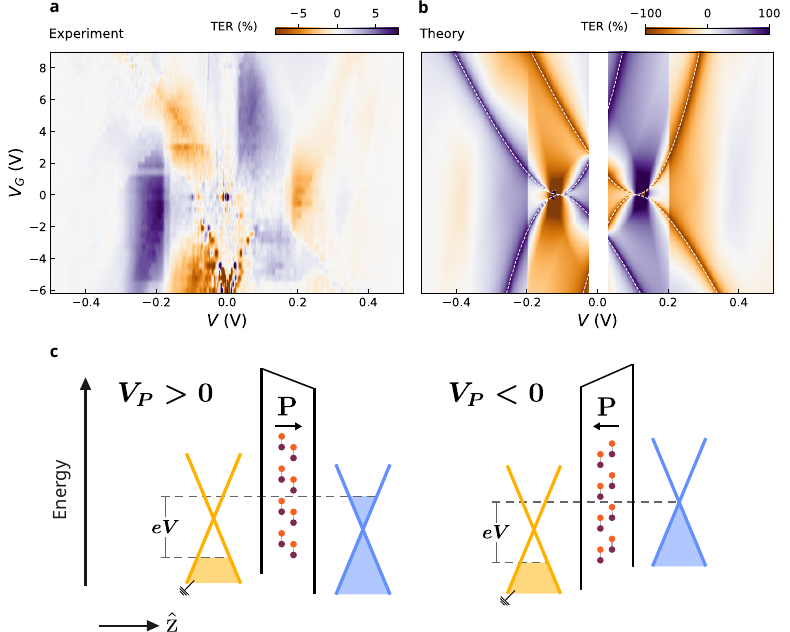}
  \caption{ \textbf{Bias and gate voltage dependence of TER. }(a) Measured TER as a function of $V$ and $V_G$. At each gate voltage a forward and backward sweep of $V$ is performed to obtain the TER. (b) TER computed from the theoretical model described in the text using Eq.\ \eqref{tunneling_integral} (see Supplementary Information Sections S3 and S4 for details of the model). For color plots in (a) and (b), TER $= 0$ is set to white. The white dashed lines trace out points of charge neutrality in one of the graphene layers for the two polarization states. (c) Illustration of underlying mechanism for observed TER. The two band diagrams describe tunneling from the right (blue, top graphene) Dirac cone to the left (yellow, bottom graphene) Dirac cone across the sliding ferroelectric barrier. Both band diagrams represent the same bias and gate voltage  (specifically, $V = -0.2$  V and $V_G = -2.5 $ V ). As the polarization switches from right, the carrier densities of the graphene layers are modified. 
  }
  \label{fig_4}
\end{figure*}

As $V_G$ is tuned, the pattern of the TER signal evolves.  For example, for $V_G = 2.2$ V (Fig.\ \ref{fig_2}(c)), the initial TER signals observed near $V = \pm 0.2$ V decrease in magnitude and two new features emerge, a negative peak at  $V = -0.11 $ V and positive peak at $V = 0.07$ V.  Upon further increasing the gate voltage to $V_G = 3.8$ V (Fig.\ \ref{fig_2}(d)), these new peaks at $V = -0.11$ V and $V = 0.07$ V overtake the the initial  two peaks from Fig.\ \ref{fig_2}(a) and (b) in magnitude. The TER for $V_G = 5.4$ V (Fig.\ \ref{fig_2}(e)) exhibits only a negative peak at negative bias, $V = -0.15$ V,  and a positive peak at positive bias,  $V = 0.07$ V -- that is, signals opposite in sign compared to $V_G = -1.0$ V (Fig. \ref{fig_2}(a)).  The full dependence of the TER on $V$ and $V_G$ is shown as a color plot in Fig.\ \ref{fig_4}(a).

\section{Tunneling model}
Although the evolution of the TER appears complicated, with multiple sign changes, this behavior can be explained qualitatively within a simple model that tracks how the electron chemical potentials in the graphene electrodes evolve as a function of changing $V$ and $V_G$.
Within this model, we calculate the Fermi energies of the top and bottom graphene layers, $E_F^T$ and $E_F^B$ (relative to the Dirac points) as a function of $V$ and $V_G$ by taking into account both the geometric capacitances within the device and the quantum capacitance of the graphene layers \cite{britnell_field-effect_2012}. The tunneling current is then computed by considering tunneling between the two Dirac cones across the ferroelectric barrier \cite{simmons_generalized_1963, wolf_principles_2012}.  Since the tunneling signal is dominated by inelastic tunneling in our devices for the range of $V$ in which the TER is large, in the main text we will consider a purely inelastic tunneling model.  Our treatment is approximate, in that we do not require conservation of the total crystal momentum of the electron and phonon system.  In the Supplementary Information we consider the corresponding TER signal for elastic tunneling and achieve qualitatively very similar results -- the physics behind the TER signals we observe arises from changes in the density of states for tunneling in the graphene electrodes rather than whether the tunneling is elastic or inelastic.

Within the approximations of our model, the inelastic tunneling current, $I_{\text{in}}$ due to an excitation with threshold energy $\hbar \omega$ can be written, for positive bias voltage and $T = 0$ K, as
\begin{equation}\label{inelastic_positive_bias}
\begin{split}
    I_{\text{in}}^+ \propto e \int \limits_{E_F^T - \frac{e\phi}{2}+ \hbar\omega}^{E_F^B + \frac{e\phi}{2}} dE \ & \rho(E-\frac{e\phi}{2}) \rho(E+\frac{e\phi}{2} -\hbar\omega) \\  &  \times \tilde{T}(E)
    H(eV-\hbar \omega)
\end{split}
\end{equation}
and at negative bias as
\begin{equation} \label{inelastic_negative_bias}
\begin{split}
    I_{\text{in}}^- \propto e \int \limits_{E_F^T - \frac{e\phi}{2}}^{E_F^B + \frac{e\phi}{2}+ \hbar\omega} dE \ &\rho(E-\frac{e\phi}{2} -\hbar\omega) \rho(E+\frac{e\phi}{2}) \\ & \times \tilde{T}(E)
    H(-eV-\hbar \omega)    
\end{split}
\end{equation}
where $e$ is the magnitude of the electron charge, $\phi$ is the electrostatic potential of the top graphene electrode relative to the bottom graphene electrode including the contributions from both the carriers on the graphene electrodes and the bound charge of the ferroelectric layer (note that $\phi$ is distinct from $V$ because of the quantum capacitances), $V_P$ is a polarization voltage arising from the bound charges of the ferroelectric layer, and $\rho(E) = \frac{2}{\pi \hbar^2 v_F^2}|E|$ gives the density of states in graphene where $\hbar$ is the reduced Planck's constant and  $v_F = 10^6$ m/s \cite{castro_neto_electronic_2009}.  $H(eV-\hbar\omega)$ is the Heaviside step function. 
The WKB transmission factor can be written in the form
\begin{equation}
    \tilde{T}(E) = \exp \left[  -\frac{2d_\text{tun}\sqrt{2m}}{\hbar} \sqrt{U_{\text{hBN}} -E}\right].
\end{equation}
In our calculations we assume $m = 0.5m_e$ for the effective mass of the hBN conduction band \cite{xu_calculation_1991} ($m_e$ is the bare electron mass), $U_{\text{hBN}} = 3$ eV \cite{watanabe_direct-bandgap_2004}, $V_P = \pm 109$ mV \cite{yasuda_stacking-engineered_2021}, and $d_\text{tun} = 0.9$ nm for the barrier thickness.  For simplicity we consider the case that initially  $E_F^T = E_F^B = 0$ when $V$, $V_G$, and $V_P$ are zero.  Supplementary sections S3 and S4 describe how we solve numerically for $E_F^T(V, V_G, V_P)$, $E_F^B(V, V_G, V_P)$ and $\phi(V, V_G, V_P)$ separately for the up and down polarization states and then compute the integral for the tunneling current. The conductance is calculated as the numerical derivative of $I$ with respect to $V$.

For illustration purposes we compute the inelastic current assuming excitation energies at $\hbar\omega = $0.025 and 0.2 eV using Eqs. \eqref{inelastic_positive_bias} and \eqref{inelastic_negative_bias} where each inelastic contribution is weighted equally. The resulting TER is shown in Fig.\ \ref{fig_4}(b). Notably, the model captures all of the large sign changes for how the TER evolves with  $V$ and $V_G$ (Fig \ref{fig_4}(a)), indicating that gate-induced changes in the tunneling density of states in the electrodes are the primary source of asymmetry within the device that allows for a non-zero TER signal. Within the model, the largest values of TER form loci as a function of $V$ and $V_G$ that correspond to the electrostatic conditions for one of the graphene layers to be at charge neutrality for one of the polarization states. Switching the polarization to or from this condition leads to the largest change in the tunneling differential conductance. These loci are traced out by dashed white lines in Fig \ref{fig_4}(b). Figure \ref{fig_4}(c) illustrates the band diagrams of the system for the two polarization states corresponding to the large value of positive TER at $V = -0.2$  V and $V_G = -2.5$ V (the two diagrams correspond to these same values of $V$ and $V_G$, with only the polarization direction switched). While our experimental data exhibit the same pattern of sign changes as in the model, the measured TER does not evolve as smoothly as a function of $V$ and $V_G$ as predicted by the model.  This could be because the polarization state may not remain fully fixed while sweeping $V$ within the measurement range, or because near charge neutrality the graphene exhibits disordered regions of electron-hole puddles rather than a complete absence of carriers at the Dirac point \cite{martin_observation_2008, xue_scanning_2011}. 

The model predicts that the gate-induced asymmetry in the tunneling density of states is capable in principle of producing quite large values of TER, because in the model the graphene density of states is assumed to go fully to zero at the Dirac point, so that one of the differential conductances ($G_\text{f}$ or $G_\text{b}$) can approach zero while the other remains large. This is unrealistic because due to the presence of disorder associated with background charges, polarization inhomogeneities, and imperfect stacking, the tunneling density of states does not go fully to zero at charge neutrality in real graphene \cite{martin_observation_2008, xue_scanning_2011}. The largest TER we measure is $7.9\%$ at $V = -1.9$ V and $V_G = -1.4$ V.  This measured TER is likely also reduced from its maximum possible value because the ferroelectric polarization likely does not switch over the entire area of the tunnel junction due to the presence of pinned domains \cite{liang_resolving_2025}.

\section{Conclusions}

In summary, our measurements demonstrate that the conductance of a ferroelectric tunnel junction with a sliding ferroelectric barrier and graphene electrodes is sensitive to the polarization state of the barrier and can generate a non-zero tunneling electroresistance (TER). The TER is tunable in sign and magnitude by varying the bias ($V$) and the gate voltage ($V_G$). The evolution of the TER with $V$ and $V_G$ (with its multiple sign changes) is captured well by a simple tunneling model which takes into account how the quantum capacitance of the graphene electrodes causes the electron chemical potentials in the electrodes to shift with $V$ and $V_G$. The resulting asymmetries in the tunneling density of states in the two electrodes provides the breaking of mirror symmetry required to obtain a non-zero value of TER.

\section*{Acknowledgments}
We thank Daniel Brandon and Maciej W. Olszewski for technical assistance.  This research was funded by the US National Science Foundation (NSF) grant DMR-2104268, and was performed in part at the Cornell NanoScale Facility, a member of the National Nanotechnology Coordinated Infrastructure (NNCI), which is supported by the NSF (grant NNCI-2025233).
K.W. and T.T. acknowledge support from the JSPS KAKENHI (Grant Numbers 21H05233 and 23H02052) , the CREST (JPMJCR24A5), JST and World Premier International Research Center Initiative (WPI), MEXT, Japan.

\section*{Data availability}
The data used in this work are available at: 10.5281/zenodo.15159564. 

\section*{Competing interests}
The authors declare no competing interests.

\bibliography{references}

\clearpage

\pagebreak
\pagenumbering{arabic}
\widetext
\begin{center}
\textbf{\large Supplementary Information: Gate-tunable electroresistance in a sliding ferroelectric tunnel junction}
\end{center}
%%%%%%%%%% Merge with supplemental materials %%%%%%%%%%
%%%%%%%%%% Prefix a "S" to all equations, figures, tables and reset the counter %%%%%%%%%%
\setcounter{equation}{0}
\setcounter{figure}{0}
\setcounter{table}{0}
\setcounter{section}{0}
\makeatletter
\renewcommand{\theequation}{S\arabic{equation}}
\renewcommand{\thefigure}{S\arabic{figure}}
\renewcommand{\thesection}{S\arabic{section}}

\section{Device fabrication}

\begin{figure*}[htbp]
  \centering
  \includegraphics[width=1\linewidth]{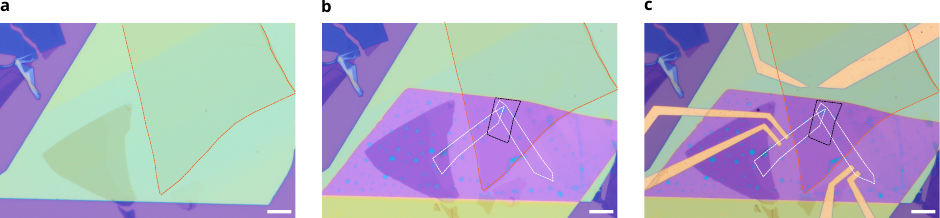}
  \caption{\textbf{Heterostructure fabrication.} (a) Bottom hBN/graphite on SiO$_2$ substrate. (b) Full heterostructure before formation of contacts. (c) Heterostructure after formation of contacts. The dashed white, black, and orange lines outline the
graphene electrodes, P-BBN, and graphite gate electrode, respectively. All scale bars are 10 $\mu$m. }
  \label{fig_s1}
\end{figure*}

Graphite and hBN are exfoliated onto 285 nm and 90 nm SiO$_2$, respectively. Graphene and monolayer hBN are identified by optical contrast. The heterostructure is assembled in two transfer steps. First, using a polycarbonate (PC) film draped over a polydimethylsiloxane (PDMS) dome \cite{zomer_fast_2014}, the bottom thick hBN (70 nm) and graphite gate electrode (2.6 nm) are picked up and deposited onto a 285 nm SiO$_2$ substrate (Fig. \ref{fig_s1}(a)). To clear an area of the exposed top surface of the hBN from polymer residue we perform atomic force microscopy in contact mode with $\sim 100 $ pN of force. The top portion of the heterostructure is similarly assembled by sequentially picking up top thick hBN (75 nm), graphene, P-BBN, graphene. The top portion is then deposited onto the bottom hBN and graphite substrate (Fig. \ref{fig_s1}(b)). The P-BBN is obtained via the tear-and-stack method \cite{kim_van_2016}. 

Once the heterostructure is assembled, we form edge contacts \cite{wang_one-dimensional_2013} (5 nm Ti/100 nm Au) to the graphene and graphite layers via electron beam lithography, reactive ion etching (CHF$_3$/O$_2$), and electron-beam metal evaporation (Fig. \ref{fig_s1}(c)). 

\section{Transport measurements}
The sample was loaded into a liquid helium dewar insert, and all  measurements are performed at $T = 4.2$  K. The DC and AC voltage biases are generated by a Keithley 2400 source and Stanford Research Systems SR860 lock-in amplifier, respectively. The AC excitation is 1 mV at 17 Hz. The device current is sent through a DL 1211 current pre-amplifier and then detected by the SR860 lock-in amplifier. Another Keithley 2400 source is used to apply the gate voltage.

\section{Electrostatic model}
We construct an electrostatic model which takes into account the quantum capacitance of the graphene layers to calculate changes in Fermi levels of the graphene layers as a function of bias voltage $V$ and gate voltage $V_G$, so that
\begin{equation}\label{elecotrstatic_barrier}
    eV = e \phi + E_F^B -E_F^T
\end{equation}
and
\begin{equation}\label{electrostatic_gate}
    eV_G = e\phi_G + E_F^B
\end{equation}
where $\phi$ is the electrostatic potential at the top graphene electrode relative to the bottom graphene electrode, $\phi_G$ is the the electrostatic potential applied to the graphite gate relative to the bottom graphene electrode, $E_F^{T (B)}$ is the Fermi energy of the top (bottom) graphene layer as measured from the Dirac point of the top (bottom) graphene layer, and $e$ is the magnitude of the electron charge. Here we assume that the quantum capacitance of the graphite gate electrode is sufficiently large that its effects can be neglected.

\begin{figure*}[htbp]
  \centering
  \includegraphics[width=0.7\linewidth]{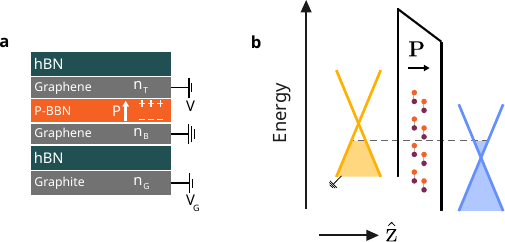}
  \caption{\textbf{Electrostatics of the heterostructure.} (a) Schematic of the device layers (not to scale) showing the relevant free carrier densities in the conducting layers and polarization from bound charges in the ferroelectric layer for $V_P > 0$ V. (b) Band diagram of the system at $V = V_G = 0$ V and $V_P> 0$ V. 
  }
  \label{fig_s2}
\end{figure*}

The electrostatic potential across the P-BBN is a superposition of the potential from the free carriers in the graphene layers and the potential from the bound charges in the P-BBN, so, 
\begin{equation}\label{electrostatic_potential}
    \phi = \frac{en_T}{C_\text{tun}} + V_P
\end{equation}
where $n_T$ is the charge density of the top graphene layer and $V_P =  \pm109$ mV is the ferroelectric potential due to the bound charges for parallel bilayer boron nitride \cite{yasuda_stacking-engineered_2021}. The first term on the right hand side of Eq.\ \eqref{electrostatic_potential} comes from Gauss's law and the charge neutrality condition $n_T + n_B + n_G = 0$. $n_B$ and $n_G$ are the charge densities of the bottom graphene and graphite gate electrode layer. For $V_P$, the positive (negative) sign is taken for the up (down) polarization state (Fig.\ \ref{fig_s2}(a)). $C_\text{tun} = \frac{\epsilon}{d_\text{tun}}$ is the capacitance per unit area of the tunnel barrier, for which we use $\epsilon = 3.4 \epsilon_0$ \cite{pierret_dielectric_2022} and $d_\text{tun} = 0.9 $ nm. The electrostatic potential of the graphite gate relative to the bottom graphene electrode will depend only on the density of free carriers and can similarly be determined, 
\begin{equation}\label{phi_eq}
    \phi_G = -\frac{e(n_T + n_B)}{C_G}
\end{equation}
where $C_G = \frac{\epsilon}{d_G}$ and $d_G \approx 70$ nm is obtained from atomic force microscopy. 

The Fermi levels of the two graphene layers can be written in terms of the carrier density in each layer as $E_F^{T(B)} = - \sqrt{\pi}\hbar v_F \text{sgn}(n_{T(B)})\sqrt{|n_{T(B)}|}$ where $\hbar$ is the reduced Planck's constant, and $v_F = 10^6$ m/s is the graphene Fermi velocity \cite{castro_neto_electronic_2009}. We treat the graphene layers as having no intrinsic doping. The electrostatic Eqs.\ \eqref{elecotrstatic_barrier} and \eqref{electrostatic_gate} can now be expressed in terms of the free charge densities and ferroelectric potential to give
\begin{equation}\label{electrostatic_barrier_n}
    eV = \frac{e^2n_T}{C_\text{tun}} +eV_P + \sqrt{\pi}\hbar v_F (\text{sgn}(n_{T})\sqrt{|n_{T}|} - \text{sgn}(n_B)\sqrt{|n_{B}|})
\end{equation}
and 
\begin{equation}\label{electrostatic_gate_n}
    eV_G = -\frac{e^2(n_T+n_B)}{C_G} + \sqrt{\pi}\hbar v_F \text{sgn}(n_B)\sqrt{|n_{B}|}.
\end{equation}
Eqs.\ \eqref{electrostatic_barrier_n} and \eqref{electrostatic_gate_n} can be solved numerically in order to determine $n_T(V, V_G, V_P)$ and $n_B(V, V_G, V_P)$ and the respective Fermi energies relative to the Dirac points.
Fig.\ \ref{fig_s2}(b) illustrates a band diagram for the system at $V = V_G = 0$ V. Due to the presence of a ferroelectric potential, the graphene layers acquire finite doping at zero bias. The two Dirac points are offset by $e\phi$ due to the total electrostatic potential.

\section{Elastic Tunneling model}

As noted in the main text, inelastic tunneling is expected to provide the dominant contribution to the current in the voltage range of our measurements for crystalline graphene/hBN/graphene tunnel junctions in which the graphene lattices are not aligned, because crystal momentum conservation does not permit elastic tunneling.  For this reason we focused on inelastic tunneling in the main text.  However, the main results of our calculation are unchanged if one uses a simpler elastic tunneling model (which would be appropriate for a ferroelectric tunnel junction with non-crytalline electrodes or barrier).

The elastic current, $I_\text{el}$, from energy conserving tunneling between two metals can be expressed as \cite{simmons_generalized_1963, wolf_principles_2012}
\begin{equation}\label{tunneling_integral}
     I_\text{el} \propto e \int dE \ \rho_B(E) \rho_T(E) [f(E) - f(E  + eV)]T(E, \phi)
\end{equation}
where $\rho_{T(B)}(E)$ is the density of states in the top (bottom layer), $f(E)$ is the Fermi-Dirac function, and $T(E, \phi)$ is the WKB transmission factor. In our case, the top and bottom Dirac cones are displaced by $e\phi$ but have the same energy dependent densities of states relative to the Dirac point, so that we have  $\rho_B(E) = \frac{2}{\pi \hbar^2 v_F^2}|E| \equiv \rho(E)$ \cite{castro_neto_electronic_2009} and  $\rho_T(E) = \rho(E+e\phi)$. 
Therefore, at $T = 0$ K, the current becomes
\begin{align}\label{tunneling_integral}
     I_\text{el} \propto  e \int \limits_{E_F^T - e\phi}^{E_F^B} dE \ \rho(E) \rho(E+e\phi) T(E, \phi)
    = e \int \limits_{E_F^T - \frac{e\phi}{2}}^{E_F^B + \frac{e\phi}{2}} dE \ \rho(E-\frac{e\phi}{2}) \rho(E+\frac{e\phi}{2}) \tilde{T}(E).
\end{align}
The WKB transmission factor is
\begin{equation}
    T(E, \phi) = \exp \left[  -\frac{2d_\text{tun}\sqrt{2m}}{\hbar} \sqrt{U_{\text{hBN}} {\color{blue}-} \frac{e\phi}{2}-E}\right]
\end{equation}
so that 
\begin{equation}
    \tilde{T}(E) = \exp \left[  -\frac{2d_\text{tun}\sqrt{2m}}{\hbar} \sqrt{U_{\text{hBN}} -E}\right].
\end{equation}

\begin{figure*}[htbp]
  \centering
  \includegraphics[width=0.5\linewidth]{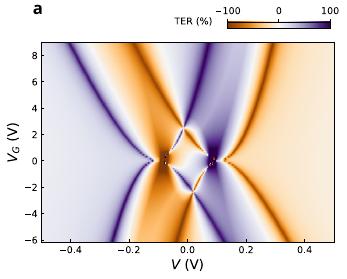}
  \caption{\textbf{Calculation of the TER signal within an elastic tunneling model.} (a) TER calculated as described in Section S4. }
  \label{fig_s3}
\end{figure*}

To compute the integral in Eq.\ \eqref{tunneling_integral}, Eqs.\ \eqref{electrostatic_barrier_n} and \eqref{electrostatic_gate_n} are first solved to obtain $E_F^T(V, V_G)$ and $E_F^B(V, V_G)$ for the up and down polarization states.  This allows the determination of $\phi$ from Eq.\ \eqref{elecotrstatic_barrier}, and then the integral for the current is computed numerically. The conductance is calculated as the numerical derivative of $I$ with respect to $V$.  The results of this calculation for the TER are shown in Fig.\ \ref{fig_s3}, where were have used the same parameters as in the main text:
 $m = 0.5m_e$ for the effective mass of the hBN conduction band \cite{xu_calculation_1991} ($m_e$ is the bare electron mass) and $U_{\text{hBN}} = 3$ eV \cite{watanabe_direct-bandgap_2004} for the energy difference between the bottom of the hBN conduction band and graphene Dirac point which we assume lies in the middle of the hBN band gap. Assuming elastic tunneling rather than inelastic tunneling produces only minor changes in the overall pattern of the TER signal.  (Compare Fig.\ \ref{fig_s3} to Fig.\ \ref{fig_4}).

\clearpage

\end{document}